\documentclass{article}
\usepackage[utf8]{inputenc}
\usepackage{graphicx}
\usepackage{subfigure}
\usepackage{algorithm}
\usepackage{amssymb}
\usepackage{amsmath}
\usepackage{amsfonts}
\usepackage{todonotes}
\title{Using Network Embeddings for Improving Network Alignment.}

\author{Pietro Hiram Guzzi}
\date{January 2020}

\begin{document}
\maketitle
\begin{abstract}

Network (or Graph) Alignment Algorithms aims to reveal structural similarities among graphs. In particular Local Network Alignment Algorithms (LNAs) finds local regions of similarity among two or more networks. Such algorithms are in general based on a set of seed nodes that are used to grow an alignment. Almost all LNAs algorithms use as seed nodes a set of vertices based on context information (e.g. a set of biologically related in biological network alignment) and this may cause a bias or a data-circularity problem. More recently, we demonstrated that the use of topological information in the choice of seed nodes may improve the quality of the alignments. We used some common approaches based on global alignment algorithms for capturing topological similarity among nodes. In parallel, it has been demonstrated that the use of network embedding methods (or representation learning), may capture the structural similarity among nodes better than other methods. Therefore we propose to use network embeddings to learn structural similarity among nodes and to use such similarity to improve LNA extendings our previous algorithms. We define a framework for LNA.

\end{abstract}{}
\maketitle

\section{Introduction}



Networks are largely used in many different fields for modelling associations among objects. For instance, in molecular biology networks model associations among genes or proteins. In social sciences, networks naturally model relation among users of social networks \cite{cannataro2010protein}.

Once that networks have been generated from real data, there is the need to compare them or to transfer knowledge from a well-known network to another. Network alignment (NA) approaches are commonly used for both aims \cite{PietroHiramGuzzi:2017bn}.
NA algorithms fall in two major categories: local (LNA) and global (GNA). LNA algorithms usually finds (small) regions of similarity among two or more networks and are used to compare small regions of similarity among two or more input networks. GNA algorithms aim to find a whole mapping among nodes, discarding local region of similarity. It has been demonstrated that GNA are best choice for transfer knowledge among networks \cite{Milenkovic2010}.

In molecular biology GNA algorithms have been used to transfer knowledge from a well-studied network to other networks, while LNA algorithms usually has been used to find corresponding substructures representing for instance protein complexes  \cite{mina2014improving,agapito2013cloud4snp}. 

Almost all LNA algorithms use as input two (or more) networks and a set of initial mapped nodes (known as \textit{seed nodes}) to build the alignment. Such information are usually derived from biological considerations, and this may cause a circularity problem or a bias \cite{mina2014improving}. Consequently, many different works tried to improve LNA algorithms using topological information as input \cite{milano2018glalign}. The rationale of these approaches is the use of topological information to quantify the structural similarity among nodes. 

More recently, it has been demonstrated that network embeddings derived methods may improve the comparison of nodes of different networks with respect to other classical methods \cite{gu2018graphlets}.

Network embeddings methods, also known as representation learning, aim to learn representation that encode structural information about the network. The idea behind these representation is to learn a mapping that embed nodes in a low-dimensional vector space preserving the similarity among nodes/substructures in the original graph.




\section{Related Work}
\label{sec:related}

Recently there is a growing interest for the introduction of novel approaches for encoding structural information about the graph, also known as \textit{representation learning} \cite{cannataro2005preprocessing,guzzi2010mu}. The idea of these approaches is to learn a mapping for nodes (or sub graphs) as points of a low-dimensional vector space \begin{math} \mathbb{R}^d \end{math} \cite{hamilton}. The goal of such methods is to realise a mapping so that geometric relationships among embedded objects reflect the structure of the original graph. After obtained embeddings may be used in other machine learning tasks (e.g. node classification) or in other graph analysis algorithms.

Despite the relatively little time in which such approaches have been introduced, there exists many algorithms and many classificaiton attempts described in some existing surveys \cite{cui2018survey,su2020network,nelson2019embed,goyal2018graph,hamilton2017representation,cannataro2013data} (citare tutti i survey). Here we follow the classification proposed by Hamilton et al by presenting a general overview and a discussion of some related methods, while the interested reader may refer to such work for a more deep discussion.

In the rest of the section we will assume that the input of representation learning algorithms is a undirected and unweighted graph $G=(V,E)$ with its associated adjacency matrix $A$ and a real-valued matrix $X$ containing node attributes  $ X \in  R ^{m x |V|}$ . The goal of each algorithm is to map each node into a vector z $\in$ \begin{math} \mathbb{R}^{d} \end{math} where $d < |V|$. 

As introduced by Hamilton et al all the embedding algorithms may be represented using two functions: two mapping functions: an encoder ($Enc:$ $V \rightarrow R^d $), which maps each node to the embedding, and a decoder $Dec$, which decodes structural information about the graph from the learned embeddings. In parallel it is possible to introduce one more function: a pairwise similarity function that quantifies the similarity among nodes (or substructure) of the graph G and a loss function that measures the difference among the similarity of two nodes and the similarity among their embeddings $l(v_i,v_j),(e_{v_i},e_{v_j})$. The simplest similarity function is the adjacency matrix, according similarity equal to 1 to adjacent nodes and equal to 0 elsewhere while a common similarity function among embeddings is the dot product between the embedding vectors.

The desired behaviour of an embedding algorithm is that given two nodes $v_i$ and $v_j$ the similarity among the embeddings should reflect the similarity among nodes, i.e. minimising the loss $l$. Formally, given two nodes $v_i$, and $v_j$ in V, their embeddings $e_{v_i}$, and $e_{v_j}$, the loss among the similarity of nodes and of the embeddings $l$, the embedding algorithm should train its paramters to minimise the overall loss  function $L=\sum_{v_i,v_j \in V}l(v_i,v_j),(e_{v_i},e_{v_j})$.

\begin{figure}[h!t]
    \centering
   \includegraphics[width=\textwidth]{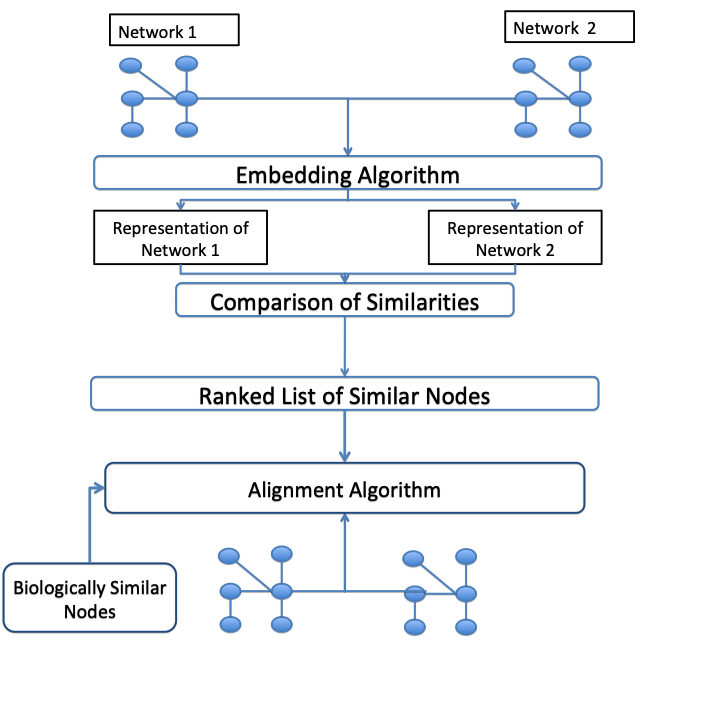}
    \label{fig:encoder decoded}
\end{figure}{}{}

\textbf{Matrix}

\paragraph{Shallow Embedding Methods}

First embedding methods rely on a simple embedding defined by Hamilton et al as \textit{shallow embedding}. In these methods each node ins encoded into a vector using a simple function as 
\begin{equation}
 ENC(v_i)=Mv_i  
\end{equation}
where M is a matrix containing the embedding vectors and $v_i$ is a vector used for selecting the column. The matrix M contains all the embeddings. Each column correspond to a node and the number of rows $d$ is lower than  the number of nodes $n$. These embeddings were initially inspired by matrix factorization approaches. The differences among these methods are in the use of different loss function and similarity measures. Examples of such methods are GraRep \cite{cao2015grarep}, Hope \cite{ou2016asymmetric}.

Main drawback of these methods is that they consider as similarity measure only the presence of edges among nodes, i.e. they consider only the first order neighbourhood of nodes. Conversely nodes may be structurally similar even they are far into the networks.
\noindent\textbf{Random Walk Based Methods}\\
Consequently a set of methods investigating higher order neighbourhoods (or local structure) have been introduced. A set of successful methods that have been introduced is based on random walks. A random walk is a random path that consists of a succession of random steps over the networks. The key idea of these methods is to derive the similarity among two nodes on the basis of the co-occurrence of nodes into random walks, i.e. two similar nodes have a great chance to co-occur in random walks. In this way the higher order neighbourhood among two nodes is encoded by the use of random walks.

The overall strategy of these methods may be summarised on three main points: \begin{enumerate}
    \item Simulation of random walks starting from each node;
    \item For each node $v$ let store the sequence of nodes visited by random walks starting from $v$;
    \item for each node $v$ learn its embedding starting from the sequence described above.
\end{enumerate}

 The difference among these methods resides on the particular definition of random walks. For instance DeepWalk uses fixed-length, unbiased random walks starting from each node \cite{perozzi2014deepwalk}. Node2Vec, instead, uses random walks having different lengths, defined as biased random walks,  \cite{grover2016node2vec}. Ribeiro et al proposed struct2vec that belong to the same class of node2vec. In struct2vec biased random walks are generated and propagated in a modified version of the network. In this network node that are structurally similar into the original network are close into distance. struct2vec learn representations for nodes in four steps:
 \begin{itemize}
     \item The input network is analysed for  determining the structural similarity between each node pair and the similarity is used to derive a hierarchy of similarity
     \item The hierarchy is used to build a weighted multilayer graph in which each layer correspond to a level of similarity
     \item Run biased random walks into the weighted multilayer graph;
     \item learn representation of the nodes from the sequence obtained by random walks.
 \end{itemize}{}

\section{Workflow of the algorithm}
\label{sec:workflow}


\subsection{Quantifying node similarity between networks.}

Given two node and their respective embeddings,  we calculate the similarity between  them using the normalised cosine similarity \cite{cannataro2010impreco}.


\begin{figure}
    \centering
    \includegraphics[width=4in]{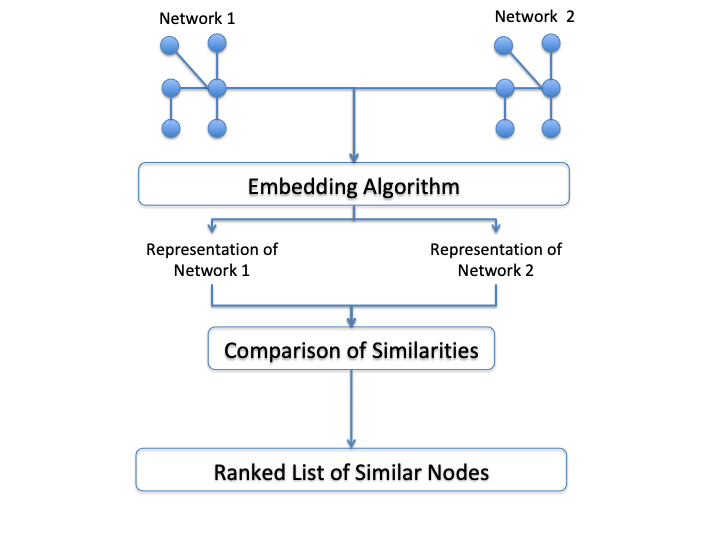}
    \caption{Quantifying Node Similarities among embeddings}
    \label{fig:embeddings}
\end{figure}

\subsection{Mixing Similarities.}

After evaluating the similarity among nodes we use as input a list of seed nodes derived from evaluating embeddings, $M_{emb}$, and the list of seed nodes derived from biological (or contextual) information Mw.
The weights of node pairs contained into $M_{emb}$ are merged with those contained into Mw following a linear combination schema producing a  final list of mapped nodes, $M_{final}$. 

\subsection{Alignment Building.}

Then we buils the local alignment using SL-GLAlign \cite{milano2018glalign}

\section{Conclusion}
\label{sec:conclusion}
It has been demonstrated that the use of network embedding methods (or representation learning), may capture the structural similarity among nodes better than other methods. Therefore we proposed to use network embeddings to learn structural similarity among nodes and to use such similarity to improve LNA extendings our previous algorithms. We defined a framework for LNA.

\bibliographystyle{plain}
\bibliography{LatestBiblio}
\end{document}